# Pair and Many-body Interactions Between Ligated Au Nanoparticles

Christopher Liepold[a], Alex Smith[a], Binhua Lin[a,b]*, Juan de Pablo[c] and Stuart A. Rice[a]*

[a]James Franck Institute, [b]Center for Advanced Radiation Sources and [c]Institute for Molecular Engineering, University of Chicago, Chicago, IL 60637

*Corresponding Authors: sarice@uchicago.edu

## Abstract

We report the results of molecular dynamics simulations of the properties of a pseudo-atom model of dodecane thiol ligated 5-nm diameter gold nanoparticles (AuNP) in vacuum as a function of ligand coverage and particle separation in three state of aggregation, namely the isolated AuNP, isolated pair of AuNPs and a square assembly of AuNPs. Our calculations show that the ligand density along a radius emanating from the core of an isolated AuNP has the same gross features for all values of the coverage; it oscillates around a constant value up to a distance along the chain corresponding to the position of the fourth pseudo-atom, then smoothly decays to zero, reflecting both the restricted conformations of the chain near the core surface and the larger numbers of conformations available further from the core. Interaction between two AuNPs generates changes in the ligand distributions of each. We examine the structure and general shape of the ligand envelope as a function of the coverage and demonstrate that the equilibrium structure of the envelope and the deformation of that envelope generated by interaction between the NPs is coverage-dependent, so that the shape, depth and position of the minimum of the potential of mean force displays a systematic dependence on the ligand coverage. We propose an accurate analytical description of the calculated potential of mean force as a function of a set of parameters that scale linearly with the ligand coverage. Noting that the conformational freedom of the ligands implies that multiparticle induced deviations from additivity of the pair potential of mean force are likely important, we define and calculate an effective pair potential of mean force for a square configuration of particles; our definition contains, implicitly, both the three- and four-particle contributions to deviation from additivity. We find that the effective pair potential of mean force in this configuration has a different minimum and a different well depth than the isolated pair potential of mean force. Previous work has found that the three-particle contribution to deviation from additivity is monotone repulsive, whereas we find that the combined three- and four-particle contributions have an attractive well, implying that the three- and four-particle contributions are of comparable magnitude but opposite sign, thereby suggesting that even higher order correction terms likely play a significant role in the behavior of assemblies of many nanoparticles.

1. Introduction

Interest in the self-assembly into stable monolayers of nanoparticles with tunable electronic, optical and magnetic properties has been driven by the opportunities to design devices that are very thin yet remarkably strong [1-9]. The constituents of such a monolayer are typically Au and semiconductor nanocrystals (cores) that are covered with organic molecule ligands, hereafter referred to as NPs. A ligand dressed nanocrystal is a very complicated many body system, so theoretical analyses of its properties, whether isolated or assembled in a monolayer, necessarily utilize simplified models of the ligand structure and the molecular interactions. A substantial number of molecular dynamics simulation studies of the ligand structure of isolated NPs and of the interactions between NPs that utilize these models have been reported [10-37]. The more sophisticated of these investigations involve simulations that use full atom [10-18] and pseudo-atom force fields [19-29], amongst which are a few investigations of the importance of the three NP interaction level deviation from additivity of the pair potential of mean force [23,24,38-40]. The results obtained from these simulations establish that the interaction between the ligands of the dressed nanoparticles, not the core-core interaction, determines the NP monolayer mechanical properties. The interpretations of the results of the simulations derived from the several approaches, taken together, are qualitatively consistent, albeit with some differences that are specific to the model representations of the NP. For the much-studied NP with Au core, the most important common features of the NP ligand structure and the NP-NP interaction obtained from the calculations can be broadly characterized as follows. First, in vacuum:

(1) The distribution of conformations of the ligands that dress an isolated Au core is sensitive to the core size and ligand density but relatively insensitive to core shape except for small (< 3 nm) cores. The angular distribution of ligands bound to a small core is much broader and more nearly uniform than that of ligands bound to a large core due to the lower maximum density of ligand dressing of the facets of a small core.

(2) For typical ligands, at the minimum of the NP-NP pair potential of mean force the core-core interaction is negligible relative to the ligand-ligand interaction. The equilibrium separation and the well depth of the NP-NP pair potential of mean force increase with the ligand length and the ligand dressing density.

(3) Because each NP is a complex system with many degrees of freedom, in a many NP assembly there are important deviations from additivity of the pair potential of mean force, so that the effective force between a pair of particles is affected by the presence of third, fourth, …, proximate particles. The available calculations show that the three-particle correction to additivity of the pair potential of mean force is an everywhere repulsive function of the NP-NP pair separation.

(4) When NPs with small cores are assembled in a two- or three-dimensional crystal, there is a threshold ligand coverage, dependent on ligand length, below which the Au cores sinter. In a three-dimensional face centered cubic crystal composed of NPs with Au cores of 2.8 nm diameter that are dressed with hexanethiol ligands that threshold is about 50% coverage; when dressed with decanethiol that threshold is about 80% coverage.

Second, when the NPs are immersed in solvent:

(5) The dry (vacuum) and wet (immersed in solvent) NP-NP pair potentials of mean force are very different. When the NP is immersed in a solvent the ligand conformations depend on the qualitative character of the ligand-solvent interaction. Using polymer chemistry terminology, a good solvent is one in which the ligand is soluble and in which it has extended conformations; a bad solvent is one in which the ligand is insoluble and in which it has compact conformations. Consequently, the wet NP pair potential of mean force is, typically, everywhere repulsive when the solvent is good and has a strong minimum when the solvent is bad.

A NP monolayer is typically dry, so its mechanical properties should be traceable to characteristic features of the ligand structure and consequent NP interactions in vacuum. The information available concerning these features has gaps that need to be filled for a better understanding of the source of the strength and other mechanical properties of a dry NP monolayer, specifically how those properties depend on the ligand structure and more information on the role played by the environment of the monolayer. Reports of the role of the environment have, to date, only concerned the case of NPs in a dense liquid solvent. However, the influence of solvent on the NP-NP interaction is perceptible on exposure to a very small amount of solvent. For example, experiments reveal that the exposure of a dry AuNP monolayer to water vapor, and its removal, generates a reversible change of the Young's modulus by almost one order of magnitude [41]. Understanding how the low concentration of water molecules at room temperature ambient vapor pressure can generate a large change in the mechanical properties of a NP monolayer requires a detailed treatment of water-ligand interactions and the consequent ligand conformation changes.

In this paper we contribute to a better understanding of the relationships between ligand conformation, ligand shell redistribution and the character of the NP pair potential of mean force in vacuum. We report the results of simulations of the interaction between model nanoparticles consisting of gold cores ligated with dodecanethiol ($CH_3(CH_2)_{11}S$) chains using a pseudo-atom representation of the ligand chains. In addition to their intrinsic interest, these simulations of the NP-NP pair potential of mean force provide a foundation for studying the importance of many particle contributions to the interaction in an assembly of NPs, a subject we address in Section 4.

The results of our simulations add to the description of the dry NP-NP interaction in the following ways:

(1) Our calculations of the interaction between two NPs with dodecanethiol dressed 5 nm Au cores include a wide range of ligand coverages; they reveal the changes in ligand conformations, the changes in position and depth of the minimum of the pair potential of mean force, and the change in shape of that minimum, as a function of ligand coverage.

(2) We develop an accurate analytic representation of the pair potential of mean force as a function of NP-NP spacing that is valid over the full range of ligand coverage that permits its calculation for arbitrary ligand coverage.

(3) We report calculations of the variation in the number density of atoms along the ligand chain as a function of ligand coverage in an isolated NP, revealing how very restricted intra-molecular atomic motion near the surface yields to much less restricted motion of the ligand chain beyond four atoms from the binding site.

(4) We show that the general shape of the distribution of ligands around the core is affected by the NP-NP interaction in different ways for different coverages. Our calculations of the number density of atoms along the ligand chain for different ligand coverages as a function of angular difference from the NP-NP centerline reveal the response of the ligand chain distribution to ligand-ligand overlap.

(5) We provide an estimate of the many NP induced deviation from additivity of the pair potential of mean force that arises from three- and four-particle proximity to a pair of NPs. Our estimate uses a different representation of the non-additivity than is commonly reported [40]. Rather than explicitly calculating the incremental third NP and fourth NP contributions to the effective force between a pair of NPs, we exploit the very limited range of the pair potential of mean force to define an effective pair potential of mean force and its variation with NP-NP separation. Our effective pair potential of mean force contains contributions from proximate third and fourth NPs. It is obtained from calculations of the total interaction free energy of a square configuration of four NP followed by division by the number of NP-NP nearest neighbors; its separation dependence is determined by varying the side length of the square. The definition clearly includes the contributions to the nearest neighbor interaction from proximate third and fourth NPs, subject to the approximation that the NP-NP interaction along the diagonal of the square is negligibly small. The latter approximation is consistent with the range of the isolated NP-NP pair potential of mean force. The calculated combined three- and four-particle interaction correction to additivity of the pair potential of mean force has a consistent dependence on ligand coverage. Accounting for these interactions generates an effective pair potential of mean force with an equilibrium spacing that increases with increasing ligand coverage but, compared with the corresponding isolated NP-NP pair potentials of mean force, has a smaller well depth when the ligand coverage is large (96%) and a greater well depth when the ligand coverage is small (32%).

(6) The three- and four-particle interaction corrections to additivity of the pair potential of mean force are of comparable magnitude, which suggests that contributions to the effective pair potential of mean force from proximity to yet more particles, e.g. fifth and sixth neighbors in a two-dimensional array, may be significant.

## 2. Simulation Model and Calculation Details

As a prelude to the description of our simulation model and details of our calculations it is worthwhile to collect some general observations. Because a single dressed nanoparticle is a complex system with many degrees of freedom, a key ingredient in the NP-NP interaction is the response of the ligand conformations to changing particle-particle separation. Calculation of the change in free energy that follows execution of a change in that separation involves averages over all conformations of the dressing ligands and, when relevant, NP rotations. Nanoparticles in a liquid suspension are free to rotate whereas NPs in a dry film are constrained to not rotate by virtue of the ligand-ligand interactions. If the ligands bound to the nanoparticle are mobile on the core surface, the free energy change that accompanies changing the separation of the NPs also involves averaging over the responsive alteration in the surface distribution of ligands. And, as already mentioned in the Introduction, since the presence of a proximate third NP will, in principle, alter all of the ligand conformations, the total free energy of interaction of three NPs will not be accurately represented as a sum of the free energies of interaction of isolated NP pairs, i.e., the total free energy of interaction of an assembly of NPs is then not pair additive. Given the complexity of conformations associated with the ligands, it is reasonable to expect further deviations from pair additivity when the free energies of four, five, …, NP configurations are evaluated.

The shape of the core of a Au nanoparticle depends on the number of atoms in that core. In general, we can think of the core as a truncated lattice with a particular shape. For small cores ($D < 5nm$), this truncated lattice is icosahedral, for moderate size cores ($5nm < D < 10nm$) it is dodecahedral, and for larger cores it takes shapes associated with various truncated fcc lattices [42,43]. However, explicit treatments of the equilibrium structure of the gold core have shown that under the temperatures and stresses required for the formation of these particles, the edges of the crystal will soften and become somewhat rounded [44]. Consequently, we expect the $5nm$ diameter cores in our model to form either an icosahedron or a dodecahedron with smoothed edges. Even within monodisperse samples, populations of both icosahedron and dodecahedron shapes are found to coexist [43]. Because of this coexistence, the smoothing of the particle edges, and the relatively small size of the lattice facets, all of our calculations use the approximation that the gold cores are spherical. Specifically, in our model the Au cores are taken to be uniform spheres constructed from identical particles that interact with a Lennard-Jones potential. An effective potential of this form has been calculated by Everaers et al. [45], which we use for the core-pseudo-atom interactions in our system; the parameters and functional form are tabulated in Tables 1 and 2 of the Appendix.

We represent the ligand molecule as a thirteen-particle chain with three types of pseudo-atoms for the $CH_2$, $CH_3$, and $S$ moieties. All pseudo-particles except those that are nearest neighbors along a chain interact through direct Lennard-Jones style potentials, while nearest neighbors interact with a harmonic bonding potential. The cis-trans structure of the chain, and the barrier to internal rotation, are characterized by describing the bending of triplets on the chain with a harmonic angular potential, and the twisting motion of quadruplets along the chain with a torsion potential. The parameters and functional forms of each of these potentials are provided in Tables 1 and 2. The specific forms we have chosen for the potentials are borrowed from Paul et al [46], who found that they describe well the experimental equation of state behavior and the local mobility of chains in a melt. Our model of the ligand molecule also retains the qualitative character of the conformations obtained from molecular mechanics (MM3) calculations more closely than do the conformations supported by the OPLS-UA and OPLS-AA force fields [27].

There is some evidence that when dressed nanoparticles such as we consider are in solution the ligands are mobile along the gold surface, whereas they are immobile in vacuum [47]. Since we are interested in the mechanical properties of a dry NP monolayer we focus attention on the NP-NP interaction in vacuum, for which case we have fixed the ligand binding sites, equidistant from one another, on the Au core. This constraint imposes an inhomogeneity on the ligand distribution and a loss of rotational symmetry of the NP. If we suppress the rotational motion of the NP, we generate a range of possible interactions between pairs of NPs. For example, if the orientation between the two NPs is such that a ligand is bound very near the center-to-center axis we expect that at small core-core separations that ligand will tilt away from the axis. Conversely, if there are bare portions of the cores along the center-to-center axis, at small separation the total interaction is dominated by the core-core interaction. Simply put, the repulsive interaction at small particle-particle separation will be strongly dependent on the relative orientation of the distributions of ligands that cap those particles. We then expect that interaction between NPs with only a few ligands will have much more variation with orientation than will that between particles that are heavily coated with ligands. We have studied this variation by carrying out simulations with different orientations of the attached ligands and simulations with different coverages of ligands. For dodecanethiol ligands on a flat substrate, the expected maximum coverage provides each ligand with an area per molecule $A = 21.3$ Å$^2$, corresponding to a surface number density $\rho = 4.7\ nm^{-2}$, which we define to be $C = 1$ [48-50]. In our calculations the coverage of the NP varied from the low value $C = 0.11$, ($\rho = 0.5\ nm^{-2}, A = 200$ Å$^2$), corresponding to a nearly bare Au core, up to $C = 0.96$, ($\rho = 4.5\ nm^{-2}, A = 22.2$ Å$^2$), corresponding to the highest coverage experimentally available.

To calculate the free energy of a pair of NPs as a function of separation we proceed as follows. For any selected fixed NP-NP separation we first carry out simulations with fixed NP-NP orientation. In these simulations the core and sulfur pseudo-atoms interact with every other particle in the system but remain stationary while the attached ligands come to conformational equilibrium around them. This constraint allows us to keep the same dynamics and topology as if the cores were free to rotate, but with the added freedom to explicitly compare differently oriented dressed NP-NP interactions. To calculate the NP-NP separation dependence of the free energy we exploit the periodic boundaries of the simulation cell as illustrated by the cartoon in Fig. 1. The direct interactions between ligand pseudo-atoms on opposite sides of the NP are very weak since their separations are of the order of the core diameter, which is an order of magnitude larger than the length scale for the interaction between the pseudo-atoms. Therefore, rather than explicitly simulating a system with two NPs in which only a small fraction of the pseudo-atoms will contribute to the direct interaction, we allow a single NP to interact with itself across a single periodic boundary. In essence, the right side of the NP will see the left side as if the ligands were attached to a different core. When we want to vary the core-core separation between the NPs we then only need to vary the size of the simulation cell. Since the core and sulfur atoms are kept with fixed position in this picture, varying the NP-NP distance in this manner samples separation without explicitly moving the particles or giving them any artificial velocity, thereby avoiding complications with changing temperature due to changes in the velocities of the pseudo-atoms. A similar method was used to probe the interaction between PbSe nanoparticles in Ref. [27].

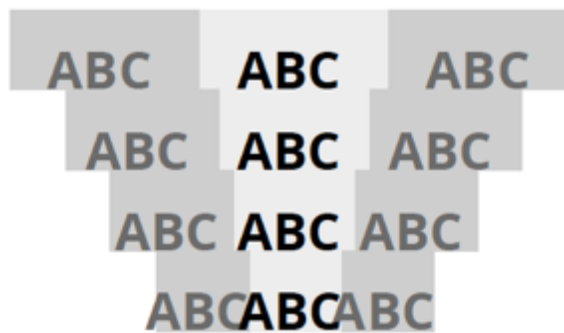

Fig. 1. Schematic representation of the periodic boundary compression method. The left side of NP "A" interacts with the right side of NP "C" across the periodic boundary. The distance between the particle centers is changed by adjusting the length of the simulation box.

All of our simulations are performed using the LAMMPS package with an NVE ensemble and a Langevin thermostat [51]. To start a simulation run we choose a random orientation for the NP and fix the simulation cell size (periodicity) so that the core-replica core separation is 100Å and the NP will not interact with its image. The NP is thermalized at 5000 K for 0.1 ns to erase memory of the ligand's initial straight-chain conformation. The temperature is then gradually reduced over another 0.1 ns until it reaches 300 K. The particle is allowed to equilibrate at this temperature for 5 ns. After equilibration, the simulation cell size is slowly reduced at a rate of 1Å/ns. Once the particle has moved 1Å, we fix the simulation cell size and let the particle equilibrate again for another 1 ns. After this equilibration we let the particle evolve for an additional 1 ns, during which time we collect data (energies and pseudo-atom trajectories). After the data collection period, the NP-NP separation is changed, and this process is repeated until the particles have reached the minimum separation of 51Å. For the region between 75Å and 51Å, we sampled with a higher spatial resolution than from 100Å to 75Å, namely every 0.33Å, changing the periodicity at 0.33Å/ns, but still equilibrating and collecting data for the same time. We repeated this procedure while varying the compression rates and equilibration times and found that allowing the system to equilibrate for substantially longer periods ($\sim 10 \times$ longer) or changing the periodicity much more slowly ($\sim 10 \times$ slower) did not substantially affect the resulting energies or pseudo-atom trajectories. Conversely, reducing the equilibration time by roughly a factor of five, or increasing the compression rate beyond 10Å/ns, led to the energy growing over the course of the simulation, implying the compression was driving the system out of equilibrium. By keeping the equilibration times long and the compression rates small, we ensure that the transitions between separations occur quasi-statically.

### 3. Results from the Calculations

**A. The Pseudo-atom Model Pair Potential of Mean Force**

For each configuration with a fixed NP-NP separation defined with respect to the image NP, fixed NP orientation, and fixed NP ligand coverage, the time average over the potential energies in the space of ligand conformations gives the constrained free energy of interaction for that NP-NP separation and NP orientation. Repetition of the procedure for different NP-NP separations yields the constrained potential of mean force for fixed NP orientation; the average of the latter over NP orientations yields the unconstrained potential of mean force. -This method of calculating the potential of mean force through constrained molecular dynamics is well described in the literature [17,24,52-57].

A sample of the results of our calculations of the pseudo-atom model NP-NP potential of mean force is displayed in Fig. 2.  Except for very low ligand coverage, in which case the shape of the NP-NP repulsive interaction is poorly constrained, we find that the pair potential of mean force separation dependence is well represented by the sum of an algebraic repulsive term, an exponential attractive term and a Gaussian correction term with coverage dependent parameters:

$$U(r) = \epsilon[(\tfrac{\sigma_R}{r})^{64} - e^{-(r-r_A)/\alpha_A} + G e^{-(r-r_G)^2/(2\sigma_G^2)}] \qquad (3.1)$$

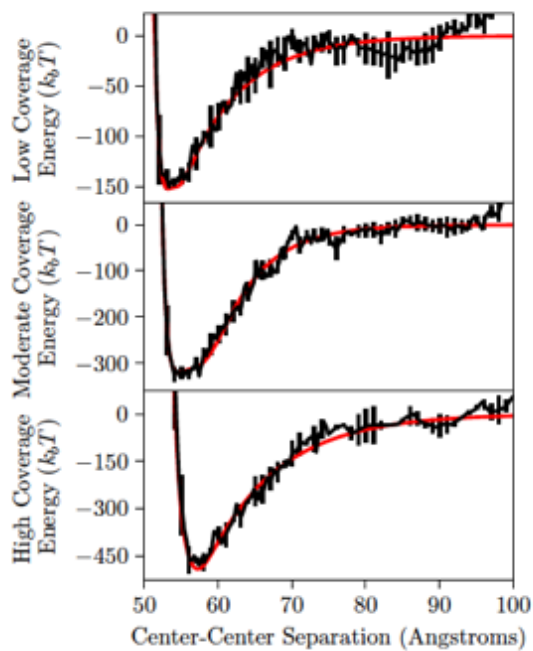

Fig. 2. Comparison of the computed pair interaction free energy (black) with the fit to Eq. (3.1).  The ligand coverages are $C_{Low}$ = 0.32, $C_{Moderate}$ = 0.64 and $C_{High}$ = 0.96.

Eq. (3.1) is found to accurately represent both the depth and the shape of the potential of mean force with variation of the ligand coverage (see Figs. 2 and 3).  As shown in Fig. 4, the length scales for the repulsion ($\sigma_R$), the well minimum ($r_0$) and the attraction ($r_A$) grow linearly with ligand coverage.

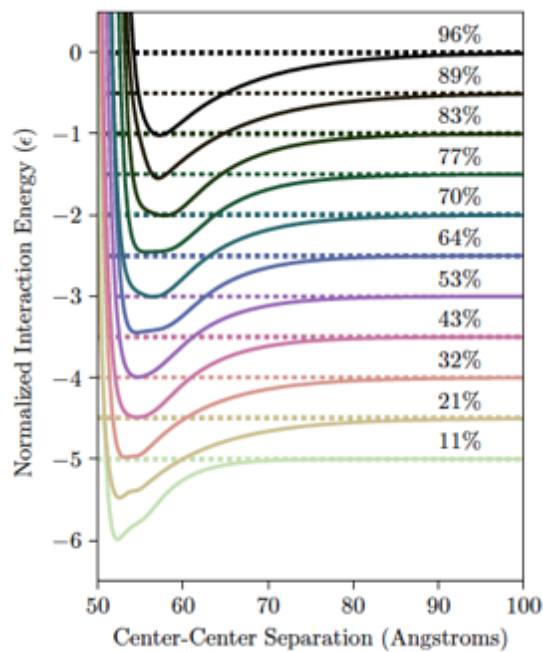

Fig. 3. The pair interaction free energies as a function of ligand coverage. The curves shown are the fits of the simulation data to Eq. (3.1). The well depths of all of the potentials have been normalized to −1 and successive potentials are shifted by 0.5 so that the shapes of the pair interaction free energies with different ligand coverages can be better compared.

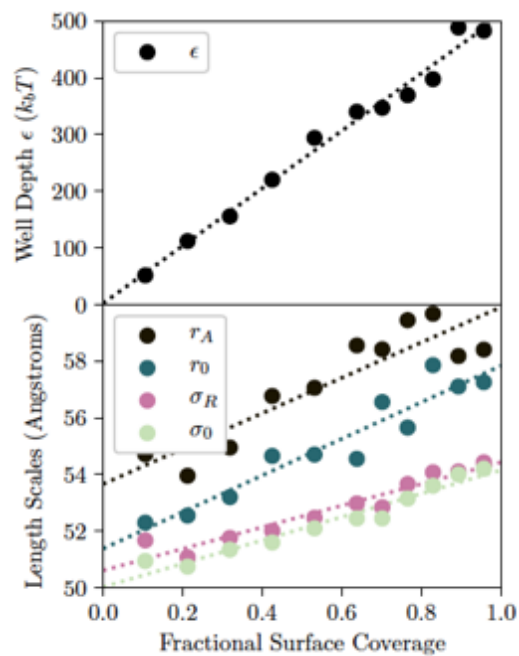

Fig. 4. The dependence of the parameters in the Eq. (3.1) on ligand coverage. The minimum of the potential is at $r_0$ and $U(\sigma_0) = 0$.

A very striking feature of the pair potentials of mean force displayed in Fig. 2 is their well depth, which grows linearly with ligand coverage up to $\sim 500 k_B T$ at the highest coverage. Similar well depths have been found in studies of other ligated nanoparticles [23,27]. Qualitative support for the reality of this finding comes from experimental studies of the structure and mechanical properties of films of dressed Au nanoparticles [58]. X-ray scattering and TEM imaging surveys of dressed Au nanoparticle films have found that the diffraction pattern and real space structure are largely independent of the surface pressure in the film, consistent with nanoparticle aggregation into large rafts even at low surface pressure, which is the expected behavior when the NP-NP attractive well depth is large compared with $k_BT$ [59,60]. And, these films have been found to be very strong, with Young's moduli on the order of several GPa, consistent with the Young's modulus of a film of NPs constructed using our pair potential of mean force [47].

These calculations reveal some interesting features of the pair potential of mean force when the ligand coverage is small. First, we find that for a few NP orientations there is negligible ligand-mediated NP-NP repulsion, so that the repulsive component of the interaction is then completely due to the core-core interaction. This situation is consistent with the occurrence of sintering, when nanoparticles bind together with their cores in contact. We only observe this behavior in our simulations at very low surface coverage, well below the coverage where it has been observed experimentally. We argue that this discrepancy can be attributed to the immobility of ligands along the surface of our simulated nanoparticles. Since a single ligand occupying the region surrounding the line of closest approach of the cores will produce a strong repulsion between the cores that arises from its bending whilst the core-core separation decreases, it will be energetically preferable for that ligand to migrate along the core surface away from the approaching core if it can. This migration will affect the distribution of core-core, core-ligand, and ligand-ligand orientations and for a specific moderate coverage will allow for more core-core interaction than does a fixed distribution of ligand coverage.

Second, at low coverage we observe that there is a large variation between the pair potentials of mean force for nanoparticles with different orientations. As described above, this variation is to be expected given the large inhomogeneity of the ligand topography in the low coverage limit. We observe that the potentials of mean force for different NP orientations become more uniform as the ligand coverage increases, and for $C$ larger than $\sim 0.43$ the variations between the different orientations sensibly vanish. In experiments, the ligand coverage is restricted by the surface chemistry between the ligands and the gold [48,61], since the surface coverage must remain in equilibrium with the concentration of ligands in the solvent surrounding the NPs. When the surface coverage is small, ligands will tend to attach to open regions on the surface, but the coverage can only remain small if the solvent is sensibly free of ligands. In the experimental preparation process the NPs are washed several times by repeatedly replacing the solvent with one that is devoid of ligands. Although the ligand coverage can be modulated, in practice it can only be reduced to about $C = 0.7$ [46]. Combining this information with the results of our calculations we infer that in the experimentally relevant situations we can treat the dressed NPs as spherically symmetric particles.

Third, the Au surface-to-Au surface separation between dodecane thiol-ligated gold NPs in a monolayer has been measured in TEM and x-ray scattering experiments [60]. The TEM data yield Au surface-to-Au surface separations in the range 1.4 nm to 1.7 nm, which is roughly the length of a single ligand molecule. The location of the minimum of the pair potential of mean force that we have calculated grows linearly with ligand coverage, from sensibly zero for nearly bare particles to 0.7 nm for fully coated particles; at comparable coverage in the simulation and experiment we find the NP-NP pair minimum to be about one third of the ligand molecule length. One source for the discrepancy between these values is the polydispersity of the gold core diameters (≈15%) used in the experimental studies, which likely weights the larger separations in the image analysis. Since the diameter of the gold core is much larger than the ligand molecule length, the fluctuations in the measured particle-particle spacing due to that polydispersity will have a comparable scale to the spacing between monodisperse particles. We believe that a more important source for the discrepancy is deviation from pair additivity of the free energy of the NP film. As will be shown in Section 4, the multi-body contributions to the NP assembly free energy are significantly large; for a ligand coverage $C$ = 0.96 the minimum of the effective pair potential of mean force including three- and four-particle contributions is about 10% larger than for the isolated pair potential of mean force, which brings the separations determined from experiment and simulation into qualitative agreement.

Fourth, the experimental finding that the measured NP-NP spacing is comparable to a chain length has led to some speculation that ligands on opposite particles may "interdigitate". We do not observe this feature in our simulation results. We find, instead, that ligands on opposite particles splay out along the mid-plane between the cores rather than mix.

## B. Pseudo-atom Model Ligand Conformations and Packing Structure

Our calculations record the positions and trajectories of the pseudo-atoms of the NPs throughout the interaction. These trajectories provide information about the internal structure of the dressing ligands, specifically the number density of pseudo-atoms throughout the ligand envelope. Since the NP pair interaction is approximately azimuthally symmetric, we define an axis linking the core-centers of the two particles. We then characterize the position of a pseudo-atom within the ligand envelope surrounding the core by its distance from the center of the core to which it is bound and the angle formed between the core-core axis and the line connecting that position to the core.

When the NP-NP pair is well separated the individual NPs are spherically symmetric except for the small-scale angular inhomogeneity mentioned above. The NP properties in this limit inform us about the equilibrium structure of the ligand envelope of an individual NP. We find that the ligand density along a radius emanating from the core, $\rho(r)$, has the same gross features for all values of C, as shown in Fig. 5.

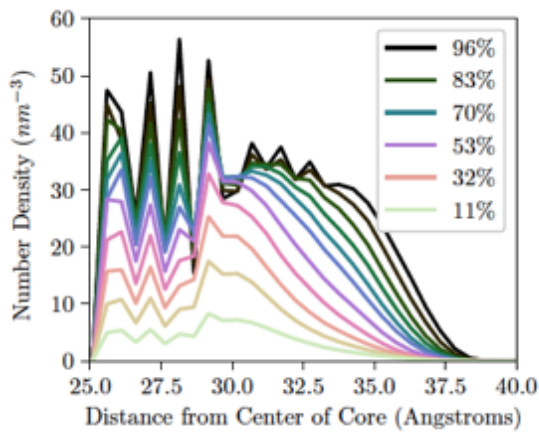

Fig. 5. Number density of pseudo-atoms as a function of distance from the NP core as a function of ligand coverage.

The pseudo-atom density close to the core oscillates around a constant value up to a distance along the chain corresponding to the position of the fourth pseudo-atom, then smoothly decays to zero. This radial structure is a consequence of the fixed positions of the ligand attachments to the NP surface. Consequently, the motions of the first few pseudo-atoms of the ligand chain are restricted. The first triplet of pseudo-atoms is confined to very nearly forming a rigid triangle, and the first quadruplet has a single *cis* and two *gauche* conformations available; these nearly fixed conformations lead to the peaks in $\rho(r)$. At greater distance from the NP surface there are substantially more conformations available to a ligand chain, including freedom to tilt away from the normal to the surface, noting that said tilt is restricted by the presence of neighboring ligands.

Near the NP core the density of pseudo-atoms in the ligand chains is proportional to the coverage on the surface, even down to very low coverage, indicating that the density in that region is strictly due to the changing number of chains attached. This observation implies that the ends of the chains are not bending back into the interior region, as that would increase the observed density at low coverage. Since the density-coverage proportionality holds down to low coverage, it cannot be attributed to exclusion effects from neighboring ligands alone. Instead, it can be understood as another consequence of the limited flexibility of the ligand chains. The torsion potential keeps the chain rigid on the scale of a third of the chain length. Although at the unbound end of the chain there are many configurations very few include chain reversal. For ligand coverage greater than *C* = 0.8 the pseudo-atom density distribution has small peaks out to about the seventh position in the ligand chain, after which it decays continuously to zero, whereas for ligand coverage less than 0.8 the pseudo-atom density distribution decays continuously to zero for separations beyond the fourth position in the ligand chain.

We now consider the ligand pseudo-atom density distribution as a function of the NP-NP separation. Snapshots of three ligand density distributions in the isolated particle and in a NP-NP pair at the respective equilibrium separations are shown in Fig. 6. Figure 7 displays the radial distributions of ligand density at several angles with respect to the center-to-center axis. At low coverage the NP-NP interaction generates a small region close to the center-to-center axis where the ligand density has been highly enhanced, but there is virtually no enhancement of the ligand density at angles beyond 15 degrees from the axis. Conversely, at high coverage the pair interaction generates only a small enhancement of the ligand density along the center-to-center axis, but that enhancement extends to angles far away from the axis. As the coverage varies from low to high there is a smooth shift between these behaviors. The fractional change in the density in the interacting region decreases gradually with increasing coverage, while the angular range over which that density is affected grows until the entire particle is involved.

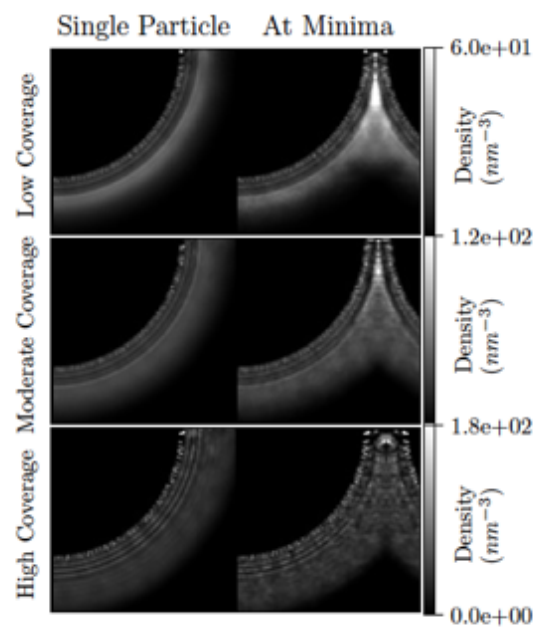

Fig. 6. Spatial distribution of pseudo-atoms around the NP core. The distributions are displayed in cylindrical coordinates after averaging over the azimuthal angle. The axis of the cylindrical coordinates coincides with the center-to center line between NPs. The left column shows the ligand distribution when the core-core separation is 10 nm; the right column shows the ligand distribution at the equilibrium core-core separation. The three frames are for ligand coverages (top-to-bottom) $C = 0.32, C = 0.64, C = 0.96$.

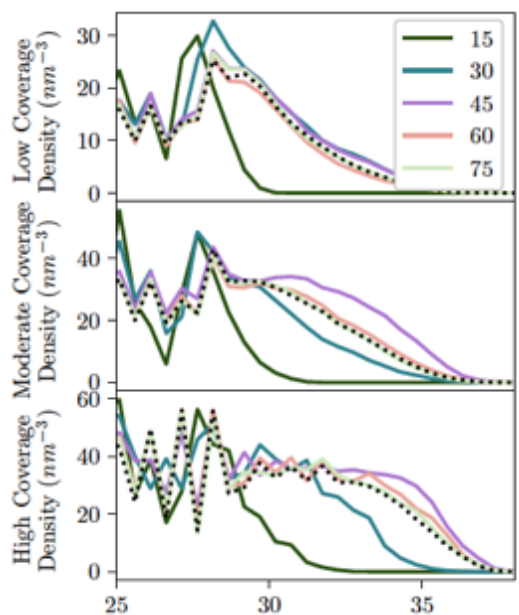

Fig. 7. Cross-sections of the spatial distributions of pseudo-atoms around the Au core along lines with fixed angular separation from the line between core centers, all for the equilibrium core-core separation for the angles labelled in the inset. The single particle ligand distributions shown in Fig. 5 are here represented as the black lines. From top-to-bottom $C = 0.32, C = 0.64, C = 0.96$.

**4. Pseudo-atom Model Non-additivity of the Pair Potential of Mean Force**

Our simulation data reproduce the qualitative features of the pair potential of mean force inferred from earlier simulations [10-37] and they extend the coverage of the parameters that influence the pair potential of mean force; the quantitative differences between the results of the several simulations arise from differences in the model NPs (e.g. core diameter and ligand coverage), and the use different parametric representations of the ligand conformations and the ligand-ligand interactions. The calculations reported in this paper clearly show that the pair potential of mean force is sensitive to the coverage of the NP by the ligands and the character of the ligand conformations, and that the distribution of ligand conformations depends on the NP-NP separation. The distribution of ligand conformations in an NP-NP pair retains statistical cylindrical symmetry, but not spherical isotropy, and the loss of the latter has been suggested to be the driver for anisotropic NP structures, such as strings of NPs [62,63]. Of more relevance to us, the distribution of ligand conformations on each of the NPs of a pair separated by some fixed distance is expected to be further changed when other NPs are nearby, which observation implies that the free energy of assembly of NPs will have a non-trivial multi-particle contribution augmenting the sum of pair potentials of mean force. Moreover, we expect that the magnitude of the deviation from pair additivity of the potential of mean force will depend on the ligand coverage.

For simple particles that interact with a central potential there exists a rarely exploited systematic formalism for calculation of the corrections to additivity of the pair potential of mean force. That formalism defines successive corrections to the representation of the triplet correlation function by the product of pair distribution functions [64,65] via an expansion in powers of the system density. However, the calculation of the coefficients in the power series is too complex to be practical for the case of the typical NP-NP interaction [66]. The calculations of the three NP correction to pair additivity of the potential of mean force that have been reported utilize a direct calculation of the work required to assemble particular configurations of three NPs in vacuum. These calculations, reported by Schapotschnikow and Vlugt [23], and by Bauer, Gribova, Lange, Holm and Gross [40], provide the following relevant information:

(1) For the dodecanethiol ligated model NPs studied by Bauer et al, when 3 NPs are in an equilateral configuration with separations $r_{12} = r_{13} = r_{23}$ equal to the two-particle equilibrium separation, the three- particle correction interaction is repulsive and amounts to about 30% of the total interaction. When the triangular configuration of particles subtends an angle greater than 60°, say with $r_{12}$ and $r_{23}$ fixed, or when $r_{12}$ is increased with $r_{23}$ and the angle between $r_{12}$ and $r_{23}$ fixed, the three-body correction interaction decreases. In general, for all the configurations considered, the correction to additivity of the pair potential of mean force is found to be sensibly monotone repulsive as a function of all pair separations.

(2) The range of the three-particle correction interaction is a fraction of a core diameter. For example, the calculations reported by Bauer et al show that for an isosceles triangle arrangement of dodecanethiol ligated cores with diameter 3.7 nm, with $r_{12}$ fixed at the equilibrium NP-NP separation determined by the pair potential of mean force ($\cong$ 1.33 core diameters), the three-particle correction interaction is reduced to sensibly zero when $r_{13} = r_{23}$ are increased to 1.7 core diameters. At fixed NP-NP separation the three-particle correction interaction is decreased when the core diameter is increased. We will make use of these observations in our treatment of the interaction free energy of a four-particle system (see below).

(3) As expected, the angular distribution of the deformation of the conformations of the ligands dressing a NP that approaches a pair of NPs with fixed separation depends on the deviation of the line of approach from perpendicular to the line of centers of the NP pair; the loss of symmetry of those deformations and overlap of their angular spreads contribute to the multiparticle correction interactions.

(4) The angular dependence of the three-particle correction interaction is strong. Using results from the simulations reported by Bauer et al, for the equilateral triangle configuration with all NP-NP separations equal to that determined by the isolated pair potential of mean force the three-particle correction interaction is approximately 250 kJ/mole, whereas with the same particle separations in the right triangle configuration the three-particle correction interaction is only 75 kJ/mole.

(5) Schapotschnikow and Vlugt have reported the results of simulations of three-particle interactions between butanethiol and octanethiol capped NPs with Au core diameter 1.8 nm. If 1, 2 and 3 identify the three NPs, they calculate an effective pair potential of mean force that is defined by integration of the projection of the sum of the forces between particles 1 and 3, and 2 and 3, onto the bisector of the separation between particles 1 and 2. With the potential parameters they used for the octanethiol capped NPs, the three-body correction to additivity of the pair potential of mean force is found to be monotone repulsive. These calculations predict that the three-particle correction to pair additivity vanishes when the NP-NP separation exceeds the distance at which the capping ligand chains overlap.

(6) For all configurations of the three particles with separations that permit ligand chain overlap it is found that the equilibrium NP-NP separation is increased from that determined by the isolated pair potential of mean force, and that the well depth is reduced from that determined by the isolated pair potential of mean force.

We now examine whether the perturbation to the capping ligand distribution of a three NP cluster that is generated by the proximity of a fourth NP is sufficiently large to generate a non-trivial correction to the additivity of the pair potential of mean force. We do not explicitly calculate the three- and four-particle correction interactions. Rather, we define an effective pair potential of mean force that plays the role of a "bond" potential of mean force; the defined potential depends only on the NP-NP separation. The definition we propose is an approximation that takes advantage of the limited range of the isolated pair potential of mean force, and of the limited range of the three-particle correction to the additivity of the potential of mean force as revealed by (i) the calculations reported in this paper that show that the pair potential of mean force of fully capped NPs at a separation of 1.4 core diameters is of order 7% of the well depth at the equilibrium separation (see Fig. 2), and (ii) the calculations of Schapotschnikow and Vlugt that show that the three-particle correction to the additivity of the potential of mean force is vanishingly small when pair separation exceeds the separation for incipient ligand chain overlap. These observations suggest that an effective pair potential of mean force that includes the aggregate three-particle and four-particle interaction corrections can be obtained by calculating total interaction free energy of an assembly of four NPs arranged in a square; the NP-NP separation of that effective pair potential of mean force is obtained from symmetry preserving expansions/contractions of the square. By construction, the total interaction free energy calculation includes whatever contributions arise from three and four NP correction interactions, hence contains hidden contributions from the ligand distribution perturbations generated by the third and fourth NPs. The omission from the definition of the effective pair potential of the interactions between NPs at the ends of the diagonal of the square configuration, with NP-NP separation $\sqrt{2}$ larger than the nearest neighbor separation, is consistent with the estimates described above of the relative magnitudes of both the pair potential of mean force and the three-particle correction to pair additivity at that separation.

We have calculated the total interaction free energies of square configurations of NPs with three different ligand coverages; Fig. 8 displays the defined effective pair potentials of mean force and Fig. 9 displays the difference between the effective potentials and the corresponding isolated pair potentials of mean force for the several ligand coverages. These calculations reveal three important features of the deviation from additivity of the pair potential of mean force. First, as reported by others, the depth of the well in the effective pair potential of mean force differs from that between an isolated pair of NPs. That difference in depth is ligand coverage dependent. Second, for the ligand coverage range considered the position of the minimum in the effective pair potential of mean force is shifted to larger separation than that between an isolated pair of NPs with the same coverage. We find that the well of the effective pair potential of mean force is deeper at small coverage and shallower at high coverage than that of the pure pair potential of mean force.

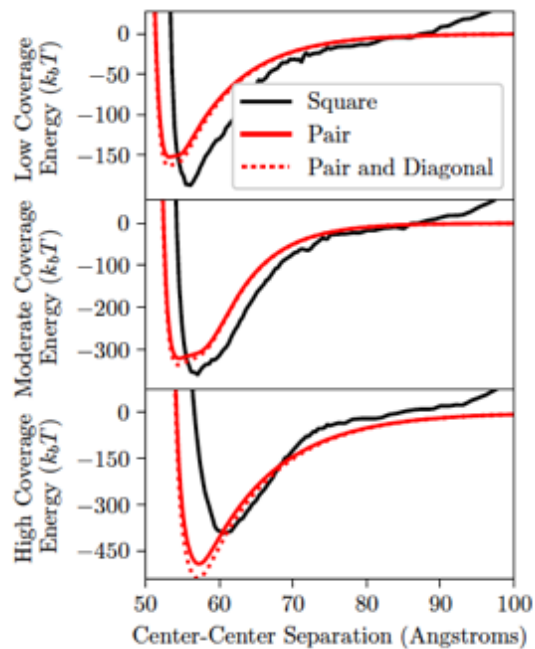

Fig. 8. Pair free energy function for an isolated NP pair (red) and inferred from the free energy of four NPs in a square (black). From top-to-bottom $C = 0.32, C = 0.64, C = 0.96$.

We note that the calculated combined three- and four-particle interaction corrections to additivity of the pair potential of mean force as a function of NP-NP separation has a consistent dependence on ligand coverage. Accounting for these interactions we find an effective pair potential of mean force (Fig. 8) with an equilibrium spacing that increases with increasing ligand coverage but, compared with the corresponding isolated NP-NP pair potentials of mean force, has a smaller well depth when the ligand coverage is large (96%) and a greater well depth when the ligand coverage is small (32%).

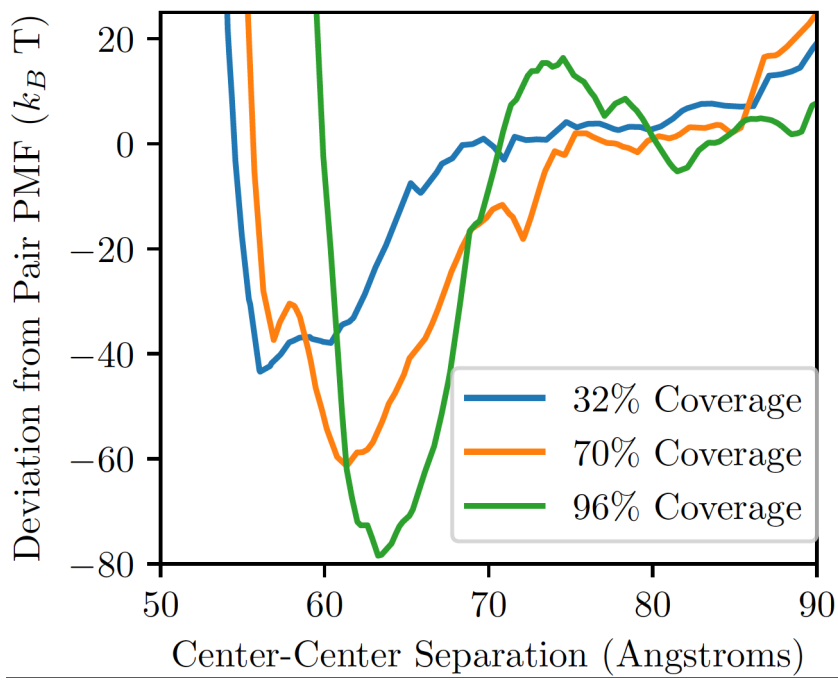

Fig. 9. Difference between the effective pair potential of mean force and the isolated pair potential of mean force. From top-to-bottom $C = 0.32, C = 0.64, C = 0.96$.

We now examine how the three- and four-particle correction interactions to the pair potential of mean force generate the observed changes in well depth and NP-NP equilibrium separation. Our definition of the effective pair potential of mean force does not reveal the partitioning of the correction interactions between the three- and four-particle contributions. The calculations reported by Schapotschnikow and Vlugt, and by Bauer et al, both show that for fully capped NPs the correction interaction induced by the third-particle perturbation of a pair of NPs is monotone repulsive. The magnitudes of the well depths displayed in Fig. 9, combined with the magnitudes of the three-particle correction interaction computed by Schapotschnikow and Vlugt, and by Bauer et al, imply that the four-particle correction interaction is comparable in magnitude to the three-particle correction interaction but of opposite sign, i.e. attractive. We argue that the importance of the many particle corrections to additivity of the potential of mean force is visualized in dramatic changes in the ligand distribution on the surface of the Au core. We show in Fig. 10 the ligand distribution on one AuNP in the square assembly when the AuNPs are far apart (right column, center-to-center separation 8 nm) and close together (left column, center-to-center separation 5.3 nm). The dramatic transition from a spherical distribution to a near square distribution with greatly enhanced density of ligands at the corners is apparent for all three coverages (1.5 ligands/nm$^2$, 3.0 ligands/nm$^2$, 4.5ligands/nm$^2$). We conclude that it is plausible that contributions to the effective pair potential of mean force from proximity to yet more particles, e.g. fifth and sixth neighbors in a two-dimensional array, may be significant.

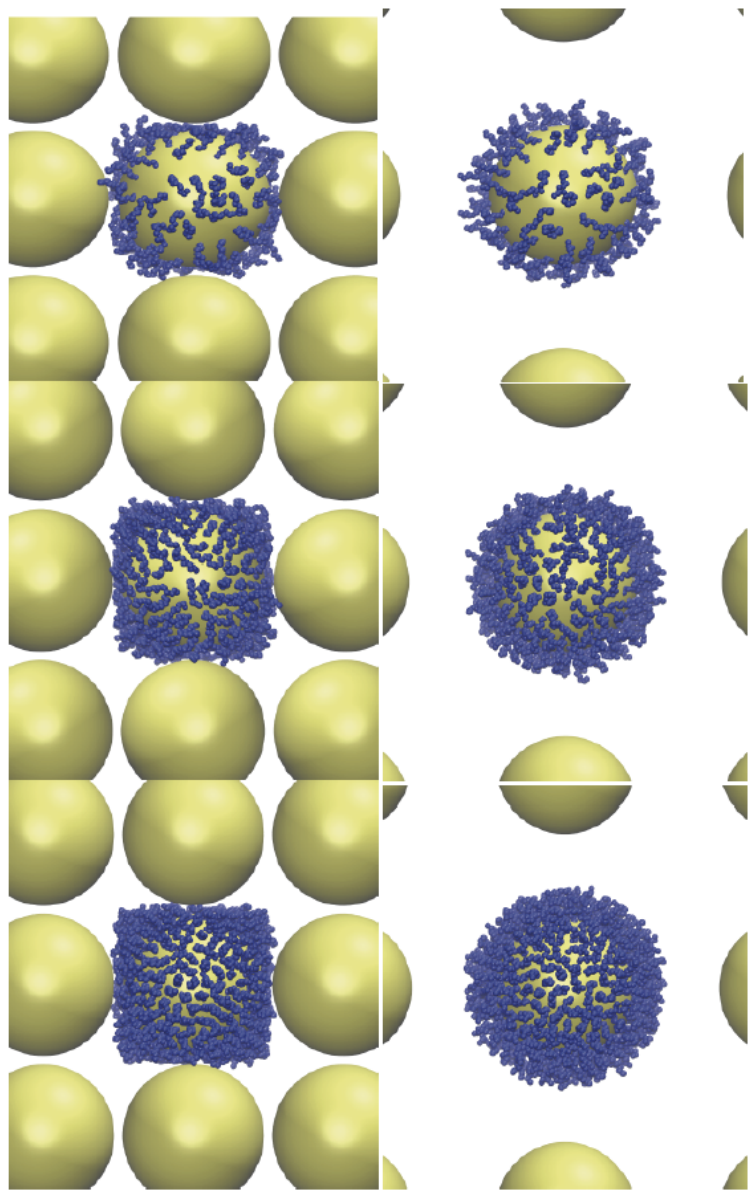

Fig. 10. Ligand distributions on a AuNP in a square array of particles: Right column, center-to-center separation 8 nm, left column, center-to-center separation 5.3 nm, top row 1.5 ligands/nm$^2$, middle row 3.0 ligands/nm$^2$, bottom row 4.5ligands/nm$^2$.

## 5. Discussion

In summary, our simulations of the properties of a pseudo-atom model of a dodecanethiol-ligated gold core nanoparticle in vacuum predict that the pseudo-atom density distribution in the interior region of the ligand envelope is structured with a scale length of the pseudo-atom-pseudo-atom separation, and that the outer region of the envelope has a pseudo-atom density profile that smoothly decays to zero. The NP-NP interaction in vacuum is strongly dependent on the ligand coverage, and therefore on the density and distribution of pseudo-atoms in the envelope. In turn, the distribution of pseudo-atoms in the ligand envelope changes as the NP-NP separation changes. We have developed a simple functional form for the pair potential of mean force; this functional form has length scales characteristic of short-range repulsion and the well location, and a well depth, all of which scale linearly with the surface coverage; the well depth which reaches up to $\sim 500 k_B T$ at the highest coverage. At the minimum of the pair potential of mean force between two NPs, when the ligand coverage is low there is localized enhancement in the density of pseudo-atoms between the particles; when the ligand coverage is high the pseudo-atom density change is spread around the entire nanoparticle. The curvature at the minimum of the pair potential of mean force is consistent with the measured Young's modulus of a monolayer of dodecanethiol-ligated gold NPs [46]. Because the internal structures associated with the ligand chains are dependent on the ligand coverage and on the NP-NP separation, there are important multi-particle contributions to the effective pair potential of mean force. Our calculations imply that the three- and four-particle correction interactions to the additivity of the pair potential of mean force are of comparable magnitude but opposite sign, and that correction interaction contributions from yet larger numbers of particles need to be investigated.

Several phenomenological models have been proposed to describe the interactions between capped nanoparticles. In particular, the Optimum Packing Model (OPM) [67] and the Overlapping Cone Model (OCM) [23] have been widely used to suggest equilibrium separations between nanoparticles. These models set the equilibrium separation as that where the volume occupied by some ligands in the system matches the free volume available for those ligands. The OPM compares the volume occupied by a single fully-extended ligand with the cone formed by the center of the core, the area of the binding site, and the midplane between the nanoparticles. The OCM compares the volume of the cone formed by the center of the core, the midplane between the nanoparticles, and the edge of the ligand envelope with the volume of the ligands which are bound within that cone. Both models are intuitively easy to grasp and predict equilibrium separations which are in good agreement with experimental data, but the microscopic picture proposed does not agree with the results of our simulations. In essence, the OCM and the OPM assume that within the overlapping region the density of particles grows until the region is fully packed.

This increase in the density fully accounts for the lost free volume around the nanoparticle, so the density does not increase outside the cone region. But our simulations contradict this feature of the models; they show that at moderate to high ligand coverages the enhanced-density regions extend well beyond the overlapping cone region.

## 6. Acknowledgements

The research reported in this paper was primarily supported by the University of Chicago Materials Research Science and Engineering Center, funded by National Science Foundation Grant No. DMR-1420709, and partially supported by a Senior Mentor Grant from the Camille and Henry Dreyfus Foundation (Grant No. SI-14-014). B.L. acknowledges support from Chem. Mat-CARS (NSF/CHE-1346572).

# Appendix

# Pseudo-atom potential parameters

**TABLE I. Interactions Used in Model**

| | |
|---|---|
| Core-Core | $U_{cc}(r) = -\dfrac{A_{cc}}{12}\left[\dfrac{D^2}{r^2 - D^2} + \dfrac{D^2}{r^2} + 2\ln\left(\dfrac{r^2 - D^2}{r^2}\right)\right]$ $+ \dfrac{A_{cc}}{75600}\dfrac{\sigma_c^6}{r}\left[\dfrac{2r^2 - 14DR + 27D^2}{(r-D)^7} + \dfrac{2r^2 + 14Dr + 27D^2}{(r+D)^7} - \dfrac{4r^2 - 30D^2}{r^7}\right]$ |
| Core-Pseudoatom | $U_{cp} = \dfrac{16 D^3 \sigma_{cp}^3 A_{cp}}{9(D^2 - 4r^2)^3}\left[1 - \dfrac{64\sigma_{cp}^6(5D^6 + 180 D^4 r^2 + 1008 D^2 r^4 + 960 r^6)}{15(D-2r)^6(D+2r)^6}\right]$ |
| Unbonded Pseudoatom-Pseudoatom | $U_{pp} = 4\epsilon\left[\dfrac{\sigma^{12}}{r^{12}} - \dfrac{\sigma^6}{r^6}\right]$ |
| 2-Pseudoatom Bond | $U_B = K_B(r - r_B)^2$ |
| 3-Pseudoatom Angle | $U_A = K_A(\theta - \theta_A)^2$ |
| 4-Pseudoatom Torsion | $U_T = A_0 + A_1 \cos(\theta) + A_2 \cos^2(\theta) + A_3 \cos^3(\theta)$ |

**TABLE II. Parameters for Interactions in Model**

| | | |
|---|---|---|
| Core-Core | | $D = 50\,\text{Å}$ |
| | | $\sigma_c = 2.934\,\text{Å}$ |
| | Au-Au | $A_{cc} = 45.0\,\text{kcal mol}^{-1}$ |
| Core-Pseudoatom | | $D = 50\,\text{Å}$ |
| | | $\sigma_{cp} = 3.47\,\text{Å}$ |
| | Au-S | $A_{cp} = 0$ |
| | Au-CH$_2$ | $A_{cp} = 24.6\,\text{kcal mol}^{-1}$ |
| | Au-CH$_3$ | $A_{cp} = 41.6\,\text{kcal mol}^{-1}$ |
| 2-Pseudoatom | | $\sigma = 4.01\,\text{Å}$ |
| | S-S | $\epsilon = 0.0934\,\text{kcal mol}^{-1}$ |
| | S-CH$_2$ | $\epsilon = 0.0934\,\text{kcal mol}^{-1}$ |
| | S-CH$_3$ | $\epsilon = 0.1455\,\text{kcal mol}^{-1}$ |
| | CH$_2$-CH$_2$ | $\epsilon = 0.0934\,\text{kcal mol}^{-1}$ |
| | CH$_2$-CH$_3$ | $\epsilon = 0.1455\,\text{kcal mol}^{-1}$ |
| | CH$_3$-CH$_3$ | $\epsilon = 0.2264\,\text{kcal mol}^{-1}$ |
| 2-Pseudoatom Bond | | $K_B = 600\,\text{kcal mol}^{-1}\,\text{Å}^{-2}$ |
| | | $r_B = 1.53\,\text{Å}$ |
| 3-Pseudoatom Angle | | $K_A = 60\,\text{kcal mol}^{-1}\,\text{degree}^{-2}$ |
| | | $\theta_A = 109.5°$ |
| 4-Pseudoatom Torsion | | $A_0 = 1.553\,\text{kcal mol}^{-1}$ |
| | | $A_1 = 4.06\,\text{kcal mol}^{-1}$ |
| | | $A_2 = 0.867\,\text{kcal mol}^{-1}$ |
| | | $A_3 = -6.48\,\text{kcal mol}^{-1}$ |